\documentstyle[12pt]{article}
\title{Squeezed Vacuum Interferometry}
\author{Matteo G. A. Paris \thanks{E-mail Address: PARIS@PV.INFN.IT} \\
{\em Istituto Nazionale di Fisica della Materia, Sezione di Pavia,} \\
{\em via Bassi 6, I-27100 Pavia, Italy}}
\date{April 20,1995}
\begin{document}
\titlepage
\maketitle
\begin{abstract}
A high-sensitive interferometric scheme is presented. It is based on
homodyne detection and squeezed vacuum phase properties.
The resulting phase sensitivity scales as $\delta\phi \simeq \frac{1}{4}
\bar{n}^{-1}$ with respect to input photons number.
\end{abstract}
\vfill
Accepted for publication in Modern Phyiscs Letters B
\newpage
The task of any interferometric measurement is to monitor fluctuations
in some environmental parameter through the detection of the induced
phase shift on the electromagnetic field. The pressure of the impinged
radiation, as a general rule, disturbs the monitored parameter itself
and therefore the problem is that of optimizing the instrumental
precision introducing into the apparatus as little energy as possible.
Interferometric measurements are widely utilized in many fields of
physics and generally represent an accurate detection, even when
classical light is placed at the input. However, in detecting
gravitational waves \cite{bondurant} or for monitoring minute variations
in the refraction index of a medium very accurate measurements are
required. Making use of nonclassical light is expected to lead to the
necessary improvement in phase sensitivity, beating shot
noise limit $\delta\phi \sim \bar{n}^{-1/2}$ with respect to the
input photons number $\bar n$.
In the last decade \cite{yurke} and especially in
the recent years \cite{invit,burnett,paris,optimized} this possibility has
been investigated and some detection schemes have been presented. Here I
suggest a high-sensitive interferometric detection scheme which is based
on highly nonclassical properties of squeezed vacuum state of radiation
field. \par
A generic interferometric scheme is outlined in Fig. \ref{f:scheme}.
The stable configuration corresponds to a fixed phase shift
$\phi_0$ between the input signal and a stable reference (local
oscillator), for example an
intense laser beam. Without loss of generality I consider
$\phi_0\equiv0$ throughout the paper. A fluctuation in the considered
parameter induces a phase shift on the input signal thus leading to a
different probability distribution of the detector outcomes. The phase
sensitivity of the whole apparatus, namely the smallest shift which can
be resolved, depends both on the apparatus itself and on the quantum state
of radiation at the input. Once the detection scheme has been chosen
sensitivity depends only on the behaviour of input signal under phase
shift evolution, in other words as it alters by the action of the
operator $\hat U = \exp\{i\hat n \phi\}$.
\par
Squeezed vacuum $|0,\zeta\rangle \equiv \hat S (\zeta )|0\rangle$ can be
obtained from vacuum state by the action of the squeezing operator
$\hat S (\zeta )=1/2 (\zeta a^{\dag2} -\bar\zeta a^2)$ with
$\zeta=re^{i\psi}$. Fluctuations of the field are increased in the
direction individuated by $\psi$ and correspondingly decreased in the
orthogonal direction. Squeezed states $|\alpha,\zeta\rangle$ are then
obtained by acting with the displacement operator $\hat D (\alpha ) =
\alpha a^{\dag} - \bar\alpha a$ on squeezed vacuum. Squeezing a state
increases its energy, in particular squeezed vacuum possesses a mean
photon numbers given by $\bar n=\sinh^2 r$.
The property of squeezed vacuum I deal with is expressed by the formula
\begin{eqnarray}
e^{i\hat n \phi}\; |0,\zeta\rangle =
|0,\zeta e^{i2\phi} \rangle
\;.\label{evol}
\end{eqnarray}
 From Eq. (\ref{evol}) two crucial informations can be extracted:
first squeezed vacuum is invariant under phase evolution and
moreover it enhances the occurred phase shift.
It is worth noticing that this is a peculiar feature of squeezed vacuum
which is lost for non-zero amplitude squeezed states. In this case, in
fact, signal and squeezing phases differently evolve
\begin{eqnarray}
e^{i\hat n \phi}\; |\alpha,\zeta\rangle =
|\alpha e^{i\phi},\zeta e^{i2\phi} \rangle
\;,\label{evol1}
\end{eqnarray}
leading to a state with broad phase distribution. The situation is
illustrated in Fig. \ref{f:shifts} where I report the Wigner
distribution function
\begin{eqnarray}
W(\alpha,\bar{\alpha}) =\int {d^2 \lambda\over\pi^2}
e^{\alpha\bar\lambda-\bar\alpha\lambda}\; \hbox{Tr}\left( \hat\rho
e^{\lambda a^{\dagger}-\bar\lambda a} \right)
\;,\label{wdef}
\end{eqnarray}
in the complex plane of the field amplitude for a squeezed vacuum and a
squeezed states along with their shifted counterparts. \par
\vskip.15cm
How the evolution (\ref{evol}) of squeezed vacuum under phase shift
can be utilized in improving interferometry ? \par
\vskip.15cm
Homodyne detector is depicted in Fig. \ref{f:homo}. The input signal is
mixed by a beam splitter with an intense laser beam $|z\rangle$
(local oscillator) and the resulting two output fields are detected by
a pair of photocounters. Homodyne output is the difference photocurrent
rescaled by the local oscillator intensity
\begin{eqnarray}
\hat x = \frac{1}{|z|} (a^{\dag}b+ab^{\dag})
\;.\label{current}
\end{eqnarray}
For sufficient intense local oscillator the probability distribution of
the homodyne photocurrent approaches \cite{grabow} that of field
quadrature $\hat a_{\varphi}=1/2(ae^{-i\varphi}+a^{\dag}e^{i\varphi})$
of the input field, $\varphi$ being the phase difference between the
signal and the local oscillator.
I consider a detection scheme in which
each experimental event consists in a couple of independent homodyne
measurements with $\varphi$ shifted by $\pi/2$ each other. This can be
easily achieved by changing the optical path of the local oscillator.
The outcomes of such an experiment can be represented as points in the
complex plane $\alpha=x+iy=\rho e^{i\phi}$ of field amplitude. It can be
shown \cite{optimized} that they are distributed according with the Wigner
function of the considered state
\begin{eqnarray}
p(x,y)=p_{\varphi} (x) p_{\varphi+ \pi/2} (y) \equiv W(x+iy,x-iy)
\;.\label{2dprob}
\end{eqnarray}
The phase value inferred from each event is the polar angle of the point
itself. The experimental histogram of the phase distribution is
obtained by dividing the plane into angular bins and then counting the
number of points which fall into each bin. In Fig. \ref{f:example} a
Monte Carlo simulation of the above experimental procedure is
illustrated for a squeezed vacuum with $\bar n = 3$.
\par
The phase probability distribution is the marginal distribution,
integrated over the radius, of the Wigner function (\ref{2dprob})
\begin{eqnarray}
p(\phi ) &=&  \int_0^{\infty} \!\! \rho d\rho\; W(\rho e^{i\phi}, \rho
e^{-i\phi}) \nonumber \\
         &=& \frac{1}{2\pi}
\frac{1}{e^{-2r}\cos^2 (\phi-\phi_0 )+e^{2r}\sin^2 (\phi-\phi_0) }
         \nonumber \\
         &\simeq& \frac{1}{2\pi}
\frac{\bar n}{1+16\bar n ^2 \sin^2 (\phi-\phi_0) }
\;.\label{pfi}
\end{eqnarray}
It exhibits sharp peaks of equal height at $\phi=\phi_0,\phi_0\pm\pi$
and deep minima (also of equal height) at
$\phi=\phi_0\pm\pi/2$ (see Fig. \ref{f:fringes}). When fluctuations in
the monitored parameter change the stationary value $\phi_0$, the phase
distribution rigidly translates as it happens for classical
interference fringes.
The presence of multiple peaks (phase bifurcation \cite{varro}) would
seem to indicate a loss of phase information, but the fact
that probability distribution rigidly translates under phase shift
reverses this assertion as the bifurcation itself can be utilized as a
multiple check for improving sensitivity.
\par
A good parameter to evaluate the phase sensitivity of the present
detection scheme is given by the full width half maximum (FWHM) of the
distribution peaks. From Eq. (\ref{pfi}) we obtain
\begin{eqnarray}
h_{1/2} \simeq \frac{\bar n}{4\pi}
\;,\label{hm}
\end{eqnarray}
for the half maximum height and correspondingly a phase sensitivity
given by
\begin{eqnarray}
\delta\phi\simeq \frac{1}{4\bar n}
\;.\label{dfi}
\end{eqnarray}
In Eq. (\ref{dfi}) is already taken into account the squeezing
enhancement described by Eq. (\ref{evol}).
\par
In conclusion, a high-sensitive interferometric scheme based on phase
properties of squeezed vacuum has been presented. It involves couples
of homodyne measurements whose outcomes are points in the complex plane
of field amplitude. The phase distribution is then obtained as marginal
probability integrated over the radius.
Multiple peaks of squeezed vacuum phase probability becomes useful
for detection as the whole distribution rigidly translates under phase
shifts. The resulting phase sensitivity is largely improved in comparison
with shot noise limit and it scales as $\delta\phi\simeq \frac{1}{4}
\bar n^{-1}$ with respect to the input photons number.
\vskip.25cm
I thank Valentina De Renzi for continuing encouragement and reading the
ma\-nu\-script. I also thank Gianni Vannini of Trieste Physics Dpt. for
hospitality and Claudio Strizzolo of I.N.F.N. for accounting in the Trieste VAX
Cluster.
\section*{References}
\begin{description}
\bibitem[1]{bondurant} R. S. Bondurant and J. H. Shapiro, Phys. Rev. A
{\bf 30}, 2548 (1984)
\bibitem[2]{yurke} B. Yurke, S. L. McCall and  J. R. Klauder,
Phys. Rev. A {\bf 33}, 4033 (1986)
\bibitem[3]{invit} G. M. D'Ariano, Int. J. Mod. Phys. B{\bf 6}, 1291 (1992).
\bibitem[4]{burnett} M. J. Holland and K Burnett,  Phys. Rev. Lett. {\bf
71}, 1355 (1993)
\bibitem[5]{paris} M. G. A. Paris, Phys. Lett. A{\bf 201}, 132 (1995).
\bibitem[6]{optimized} G. M. D'Ariano, C Macchiavello and M. G. A. Paris,
Phys. Lett. A, {\bf 195}, 286 (1994).
\bibitem[7]{grabow} W. Vogel, J. Grabow Phys. Rev. {\bf A47}, 4227 (1993).
\bibitem[8]{varro} W. Schleich, R. J. Horowicz, S. Varro,
Phys. Rev. A{\bf 40}, 7405 (1989).
\end{description}
\newpage
\section*{Caption to figures}
\begin{figure}[h]
\caption{Outline of a generic interferometric scheme. Stable
configuration corresponds to a fixed phase shift between the signal and
the local oscillator, which, in turn, leads to some distribution
outcomes. Fluctuations in the parameter under examination change the
optical path of the input signal leading to a different distribution for
the detector outcomes.}\label{f:scheme}
\end{figure}
\begin{figure}[h]
\caption{Wigner distribution function $W(\alpha,\bar{\alpha})$ for
shifted squeezed vacuum and squeezed states. In (a) a squeezed vacuum
with $\bar n =4$ initially withe zero squeezing phase is subjected to a
$\pi/8$ shift and the resulting squeezed vacuum show a squeezing phase
equal to $\psi=\pi/4$. In (b) a squeezed states with total number of
photons $\bar n=4$ and $|\alpha|=1$ is initially prepared with equal
squeezing and signal phase in order to minimize phase fluctuations.
After the same shift occurred in (a) squeezing phase is twice than the
signal one and phase distribution is broadened.}\label{f:shifts}
\end{figure}
\begin{figure}[h]
\caption{Outline of homodyne detector. The input signal is represented
by the mode $a$ whereas $b$ support a highly excited coherent state,
which provides the local oscillator. Optical path of $LO$ can be tuned in
order to switch the phase difference between the two modes from $0$ to
$\pi/2$. }\label{f:homo}
\end{figure}
\begin{figure}[h]
\caption{Monte Carlo simulation of the suggested experimental procedure
on a squeezed vacuum with $\bar n=3$. The outcomes distribution
in the complex plane and the resulting marginal phase distribution are
reported.}\label{f:example}
\end{figure}
\begin{figure}[h]
\caption{Phase probability distribution for the input squeezed vacuum
and for the squeezed vacuum after than a $\pi/8$ shift is occurred. The
plots are for a state with $\bar n=3$.}\label{f:fringes}
\end{figure}
\end{document}